# Low entropy in graphene through the Co-C system

Giampiero Amato[1a], Federico Beccaria[2], Umberto Vignolo[2], and Federica Celegato[1b]

[1a] Quantum Research Laboratory, INRIM, Strada delle Cacce 91, Torino, Italy

[1b] Nanoscience and Materials Division, INRIM, Strada delle Cacce 91, Torino, Italy

[2] Department of Physics, University of Torino, via Pietro Giuria 1, 10125 Torino, Italy

**Abstract**

Uniform, mostly single-layer graphene with enhanced stability is demonstrated over Co film. The polycrystalline Co film deposited on a $SiO_2$/Si substrate gives a continuous graphene layer that is easily transferred without the aid of any polymeric support, but preserving the material quality, as evidenced by Raman analysis. Great stability to the damaging action of the laser beam, as compared to the Cu-grown material is also observed. The better structural and electrical properties of the material are interpreted in terms of thermodynamics of the cooling-down process. It is suggested that the reduction in entropy, due annihilation of vacancies caused by C atoms precipitating during cooling, directly depends on the activation energy of C solubility into Co, which is considerably high, due to Co magnetic ordering at the process temperature. Our work expands the possibility of synthesizing single-layer graphene keeping into account the thermodynamics of various C-metal systems.

**Keywords:** graphene; vapor deposition; entropy; magnetically ordered materials

---

[1a] Corresponding author: *g.amato@inrim.it*

# 1. Introduction

Chemical Vapor Deposition (CVD) on catalyzing metals promises to be a simple, reliable and relatively cheap technique for producing large areas of good quality graphene, to be employed in the future electronics generation.

Nevertheless, the great majority of reports point out that electron mobility of such systems is limited by vacancies, grain boundaries, and distorted bonds even at the nanoscale [1].

Departures from the perfect hexagonal lattice of graphene imply that atoms with vibration frequencies different from those of the perfect crystal are present in noticeable quantity. This has consequences on the entropy of the system.

Comparisons of Raman spectra carried out in CVD and exfoliated graphene samples can yield detailed information on the respective content of entropy of these systems. Exfoliated graphene is the real system as close as possible to a condition of *global* equilibrium, whereas CVD one is a paradigm of a system at *local* equilibrium. In other words, an appreciable increase of entropy contributes to minimize the Gibbs' free energy of the graphene layer, **G=H-TS** (where **H** is the enthalpy, **T** the absolute temperature, and **S** the entropy).

The first Raman feature (1590 $cm^{-1}$) of graphene is the well-known *G* peak, characteristic for $sp^2$-hybridized carbon-carbon bonds. The second prominent feature (≈2700 $cm^{-1}$) originates from a double-resonance process, which creates an electron-hole pair that recombines after two inelastic-scattering events involving phonons with opposite momenta ad is labeled as *2D* peak. If defects are present, one of the two scattering events can occur elastically and the *D* peak, observed in this case, exhibits only half the Raman shift (1350 $cm^{-1}$). Saturated, but distorted bonds can be present even in case the *D* peak is not detected. These sources of entropy have been recently revealed [1] by the observation of the *2D* peak. In fact, *2D* line width contains valuable information on bond-length variations in graphene on scales far below the laser spot size, that is, on the nanometer-scale.

Conversely, the detection of the *D* peak implies either a number of unsaturated bonds, or a non-negligible amount of atoms, weakly bound to the neighboring ones. Those weakened bonds are prone to be broken, e.g. by the same laser light employed for Raman investigation. As it has been recently reported [2], CVD graphene grown on Cu shows degradation with light exposure time and the rate of such fragmentation is dependent on the material quality at the laser spot. In terms of the Gibbs' energy, this means that the system reaches another *local* minimum of **G**, at which the entropy contribution is further increased.

A similar phenomenon has been also observed in exfoliated graphene but at laser power densities one order of magnitude higher [3], confirming a better stability of this *nearly perfect* system.

The most used catalyzing metals employed in the CVD of graphene are Ni and Cu. Copper has the lowest affinity to carbon, in fact it does not form any carbide phases [4], and has very low carbon solubility compared to Co and Ni. The low reactivity with carbon of Cu can be attributed to the symmetrical electron distribution of the 3d-electron shell $\{[Ar]3d^{10}4s^1\}$ which minimizes reciprocal repulsions. The $3d^7$ and $3d^8$ orbitals of Co and Ni are between the most unstable electronic configuration (Fe) and the most stable one (Cu). Basing on this, several investigators considered Cu and Ni as the most suitable catalysts for graphitic carbon formation, whereas Co received less attention[5-7].

Tracking carbon during the growth process by means of isotope labeling in conjunction with Raman spectroscopic mapping [8,9], demonstrated the different kinetic behavior of CVD growth of graphene on Ni and Cu. The two mechanisms of graphene growth observed on Ni and Cu can be understood in terms of the C-metal binary phase diagram, the most important difference being that, thanks to the much lower solubility in Cu respect to Ni, only a small amount of carbon can be dissolved in Cu.

Continuously imaging the C monolayer coverage on Cu using Low-Energy Electron Microscopy confirmed no C precipitation or island growth during cooling [10], suggesting that the process is confined to the surface [11].

In contrast, Co and Ni can dissolve much more C atoms. The graphene growth comes also from the precipitation during the cool-down of the process and "polygraphene" was detected in most cases [11], as a consequence of an excess of precipitation of C atoms. For Co and Ni, the solubility and precipitation process must be controlled to some extent with the annealing, isothermal growing and cooling rates.

This means that cooling is an additional way to reduce the entropy of the system. In fact, entropy related to vacancies in Cu-grown graphene can be reduced during cooling through their diffusion, solely. It can be minimized if, e.g., two vacancies collapse into a di-vacancy (the number of *loose atoms* reduce from 6 to 4) or by their motion towards the grain boundaries, which act as *vacancy sinks*. In the case of Co and Ni, on the other hand, an extra, effective mechanism for vacancy reduction is provided by the C atoms precipitating from the substrate.

The aim of this work is to elucidate the effect of the supply of C atoms during cooling on the reduction of the entropy of the growing graphene layer. For the purpose, we will discuss in detail the properties of graphene grown onto Co substrates and discuss the effect of the activation energy of C solubility. It will be shown that this parameter plays an important role in dictating the rate of C precipitation, with important benefits on the entropy of the resulting graphene film.

## 2. Experimental

Graphene was synthesized via CVD, using $CH_4$ as a precursor gas in a vacuum chamber of a Rapid Thermal Annealing apparatus [12]. For the purpose, 500 nm thick, Co polycrystalline thin films were deposited by RF Sputtering onto Si substrates with thermal $SiO_2$ (500 nm thick). The Co deposition parameters were fixed at 100 W RF power and $10^{-2}$ mbar $Ar^+$ pressure, leading to a deposition rate around 3 Å/s. The $Co/SiO_2$ catalyst was sonicated in acetone and ethanol and then inserted in the reactor chamber. After removal of the residual Co oxide in $H_2$ atmosphere for 5 min, 10 sccm of $CH_4$ and 20 sccm of $H_2$ were delivered for 5 min at a constant temperature of 1000 °C and pressure of 6.7 mbar. Cooling down was carried out in two steps: from 1000°C to 600°C in Ar and $H_2$ atmosphere at a rate of 3.5°C/s, and from 600°C to room temperature at the same rate with a flow of $N_2$ (200 sccm). Taking advantage from its high structural integrity, the resulting graphene was then transferred onto a Si substrate with 300 nm $SiO_2$, without the aid of any supporting layer. Removal of the Co film was achieved by sample dipping in a 1 M $FeCl_3$ solution. After repeated rinsing in $H_2O$, the graphene sheet was picked up with the $Si/SiO_2$ target substrate and then dried.

Differently from Cu, Co does not emit any photoluminescence when exposed to green light. This allows for accurate Raman comparisons onto the growth substrate and the destination one. A typical one, reported in Fig. 1, indicates that the graphene layer excellently withstands the transfer step, despite of the absence of any supporting material.

In order to measure the sheet resistance of the graphene sample, Cr/Au (20 and 75 nm thick, respectively) strip contacts were deposited in a E-Gun Evaporator and defined by a mechanical mask, thus avoiding any lithography technique to get rid of the contamination by PMMA and related chemicals. The metallic strips, 150 μm wide, 766 μm spaced, define three graphene areas of interest of about 1,32 $mm^2$ (see Fig. 2) The electrical measurements were performed in a Janis Cryostat [13], at room temperature and 2 10$^{-2}$ mbar pressure.

Micro-Raman analysis has been has been performed in a homemade system described elsewhere [2]. All acquisitions of Raman spectra were performed with a laser ($\lambda$ = 532 nm, wavelength) spot diameter of 1μm and power of $P \approx$ 2mW. The spectra where collected at power density $\Psi \approx 6.4 \times 10^4$ W/cm$^2$ and acquisition time $t_{acq}$ = 60 s.

## 3. Results and Discussion

SEM and AFM images of the graphene/Co system, displayed in Fig 3, show the presence of micrometer-sized holes in the Co thin film, upon which, suspended graphene can be observed. The AFM image also shows that the suspended graphene is stretched and corrugated. Even if the detailed discussion of the formation and properties of such *graphene-on-nothing* will be presented in a forthcoming paper, it will be however shown in section 4 that the occurrence of such a growth is consistent with the predictions of our model.

The SEM micrograph of the transferred sample in Fig. 3 shows the uniformity of the graphene sheet, with little bilayer islands, further confirmed by the Raman spectroscopy analysis whose results are displayed in Fig. 5. $I_D/I_G$ and $I_{2D}/I_G$ maps (50μm x 50μm) are shown. These maps confirm the uniformity of the sample, its mostly monolayer character, and the low quantity of defects. A more specific and quantifiable measure of these two features, is provided by the histograms counting the number of spots with different $I_D/I_G$ and $I_{2D}/I_G$ ratios, peaking at 0.05 and 3.4 respectively.

As described in section 2, electrical measurement has been performed on a mm-sized graphene sheet, obtaining a value of the order of 200 Ω/Sq, comparable with the results reported for high quality CVD graphene [14-17]. The resistance of the Cr-Au contacts has been evaluated to be about 30 Ω. These results highlight the advantage of getting rid of any polymer, commonly employed in the standard transfer technique and in the lithography steps, since it can affect the electrical properties of graphene, e.g. by doping. Moreover, incomplete removal of the masking polymer can degrade the quality of the contacts on graphene [18].

The possibility of transferring the graphene sheets without the aid of a supporting layer deserves attention. Defects are commonly expected to originate from unsaturated bonds at grain boundaries, affecting both mechanical stability and carrier transport. In our case it seems that there are stiffer interconnections between grains, maybe due to the high crystallographic order of the layer, even if the presence of bilayer graphene at grain boundaries cannot be ruled out. In any case, the combination of the two effects allows for a polymer-free transfer method, which preserves the electrical properties of graphene.

We analyzed the evolution of the Raman spectra in order to determine the kinetics of defect formation. An example is reported in Fig. 6, obtained by focusing the laser onto three different spots of the sample, exhibiting similar $I_D/I_G$ ratios in the pristine state, but different $I_{2D}/I_G$. The laser continuously irradiated those spots for 90 minutes and spectra were acquired at 10 min intervals.

The reported results give evidence that in Co-grown graphene, no additional defects are induced by laser irradiation, a very different behavior if compared to CVD growth on Cu foils. It has to be pointed out that the model for the kinetics of defect formation under laser irradiation proposed in a previous work [2], predicts that fragmentation preferentially occurs at defective regions. The basic assumption of the model is that at a defective region, the energy released from the non-radiative recombination of photogenerated electron-hole pairs is sufficient to break a number of weakened bonds of *loose atoms* surrounding the defect. The Co-grown samples do not display any increase of *D*-peak intensity the with exposure time, according to the negligible pristine $I_D/I_G$ ratios if compared with the CVD graphene grown on Cu foil. Then, as mentioned in section 1, graphene films with lower entropy content are more stable.

We will demonstrate that the content of entropy of the graphene films is a direct consequence of the growth mechanism. As mentioned in section 1, graphene can grow by CVD on catalyzing metals by means of two main mechanisms, the first being the isothermal growth when the substrate is exposed to the C gaseous precursor, the second is the C precipitation upon cooling-down. In case of Cu the first mechanism dominates since only a negligible amount of additional C atoms precipitate on the metal surface [10,11], whereas, in case of high C solubility, graphene forms in both ways [19].

## 4. Theory

The theoretical treatment of the problem considers the variation of the free energy of the graphene layer during sample cooling-down. Fig. 7 displays the dependence of C solubility on the inverse T. The thermally activated behavior holds for the three metals into consideration, with the following activation energy values: $\Delta_{Co}$=0.791 eV, $\Delta_{Ni,Cu}$=0.484 eV.

The peculiar value for the activation energy of C segregation in Co has been explained by Hasebe *et al.* [20] in terms of an additional contribution to the Gibbs' energy of the internal magnetic field. It is worth mentioning that graphene growth is normally carried out at temperature values below the Curie temperature of Co, even at high concentration of segregated C [20].

The total free energy ***G*** of a graphene layer in which the vacancies left by the incomplete, isothermal growth have to be annihilated by the successive precipitation step, can be written, following Hayes and Stoneham [21], as:

$$G = G_0 + nG_1 + G_2 - nG_{-1}, \qquad (1.1)$$

where ***n*** is the number of vacancies, $G_0$ is the free energy of the perfect crystal, $G_1$ the energy needed to generate one vacancy, $G_2 = -T \cdot S_c$ ($S_c$ being the configurational entropy of ***n*** vacancies), and $G_{-1}$, the energy for a vacancy annihilation.

Imposing the equilibrium condition ($\delta G/\delta n=0$) yields:

$$T\frac{\delta S_c}{\delta n} = G_1 - G_{-1} . \tag{1.2}$$

The configurational entropy $S_c$ is generally calculated by means of the Boltzmann's famous formula $S_c=k_B \ln P$ with $P$ the number of possible microstates for the same macrostate. Following Ref[21], eq. (1.2) becomes:

$$k_B T \ln \frac{N}{n} = \Delta H - T\Delta S, \tag{1.3}$$

where $N$ is the number of sites, say, the double of the areal density of primitive unit cells in graphene (1.908 $10^{15}$ cm$^{-2}$) and the difference $G_1$-$G_{-1}$ has been made explicit in terms of the enthalpy and entropy variation. By defining $c_v=n/N$, the relative vacancies concentration, eq (1.3), yields:

$$c_v = \alpha \exp(\Delta S/k_B) \exp(-\Delta H/k_B T), \tag{1.4}$$

where an additional entropy term $\alpha$ has been introduced, as additional degrees of freedom for the combinatorics needed to calculate the entropy. Nowadays, the molar enthalpy of formation of graphene is still unknown. Herein, it is estimated to be 513 kJ/mol from the bond enthalpy of graphene [22]. Consequently, this value is the same for both formation and annihilation of a vacancy, but with opposite signs, that have already been considered in eq (1.1). It derives a substantial simplification of the (1.4), which becomes:

$$c_v = \alpha \exp(\Delta S/k_B) \tag{1.5}$$

meaning that the relative concentration of vacancies decreases with a negative variation for the entropy of the graphene layer, as expected.

Relationship (1.5) has been extracted in the case of $n \ll N$, which does not hold for a largely incomplete graphene layer, like at at the beginning of the cooling stage. For this reason, eq (1.5) better describes the situation of the Cu growth, which is almost completed after the isothermal stage. The more general relationship is:

$$c_v = [\alpha^{-1} \exp(-\Delta S/k_B)+1]^{-1}, \tag{1.6}$$

but we will still keep the (1.5) for simplicity purposes, and provide the final result obtained by considering eq. (1.6).

The additional C atoms precipitating from the metal substrate will decrease the concentration of vacancies in the graphene layer according to the *T*-dependence of C solubility in the metal:

$$c_v = \beta \exp(-\Delta/k_B T). \tag{1.7}$$

To elucidate this point, we can consider an ideal inverse process in which, starting from a complete graphene layer at room *T*, a number of vacancy sites are progressively left by an equal number of C atoms *diffusing* into metal with increasing *T*. In this ideal situation, then, the parameter *β* should be the pre-exponential factor of the *T*-dependence of C solubility (see Fig.7).

Things are more complicated, because *β* can have a *T*-dependence, due to additional effects, like macro-segregation (e.g. the non uniform cooling rate of metal surfaces respect to bulk), and micro-segregation at Co grain boundaries, which can greatly affect the C atoms precipitation at the growing surface. Additionally, a further dependence of *β* on the cooling rate cannot be excluded. The intrinsic difficulty of theoretically describing the precipitation of foreign atoms [23] in phase diagrams is commonly circumvented by investigators involved in the growth of graphene by a trial-and-error approach. All the aspects related to C segregation are beyond the scope of the present paper, which is to elucidate the effect of the activation energy of C solubility on the final entropy of the graphene layer. Then, we will simply consider *β* as a phenomenological constant, reasonably lower than the pre-exponential factor of C solubility, and related to the availability of C atoms dissolved into substrate.

By comparing eqs (1.5) and (1.7), we obtain the relationship:

$$\Delta S = -(k_B \ln(\alpha/\beta) + \Delta/T), \tag{1.8}$$

or, in case of the (1.6):

$$\Delta S = -k_B \ln [(\alpha/\beta)\exp(\Delta/k_B T) - 1]. \tag{1.9}$$

Interestingly, eqs. (1.8) and (1.9) give the *entropy gain* (=reduction) when one vacancy is annihilated, by the contribution of two terms: the first in eq. (1.8) is reminiscent of the Boltzmann formula and considers the entropy of a system with $\alpha$ vacancies and $\beta$ atoms precipitating from the substrate, the second term is related to the *rate* of precipitation which is given by the activation energy of solubility. It is evident that in the normal situation in which $\beta<\alpha$, both terms are positive in sign and the total entropy variation $\Delta S$ is negative, as expected from the annihilation of one vacancy. The situation should change if an excess of atoms is provided by the substrate, which is, $\beta>\alpha$. Here, due to the larger availability of C atoms respect to vecancies in graphene. entropy could increase, e.g. in case of formation of interstitial C, or, as it is often observed [11]. in case of enhanced precipitation at metal grain boundaries, with the consequent growth of additional graphene layers.

However, larger the energy $\Delta$, lower the *entropy loss* (=augment) when the number of atoms exceeds the number of vacancies. In other words, the activation energy of C solubility $\Delta$, increases the *entropy gain* during the growth of a single graphene layer, and reduces the *entropy loss* when C atoms are in excess. This last consideration can be understood when looking at the plots of the T-dependence of the C solubility shown in Fig. 7. A crossing point between the curves of Co and Ni is found at T ~ 950 °C. This means that Co provides most of C atoms to the growing graphene layer in the first 50 °C of cooling, from 1000 °C to 950 °C. This can be viewed as equivalent to a much faster cooling-down process, similar to the common recipe for preventing the excess of precipitation or to more sophisticated approaches, like the introduction of carbon diffusion barriers in the metal film [24].

A few words must be spent to comment the Cu case. The situation described by eq. (1.8) seems not to distinguish between Cu and Ni, the $\Delta$ values being approximately equal. It must be recalled, however, that eq. (1.8) gives the *entropy gain* $\Delta S$ when one C atom is added to the forming graphene. The total amount of the *entropy gain* is obtained by multiplying $\Delta S$ times the number of precipitated atoms. This last number is dramatically lower in Cu, then, no noticeable *entropy gain* takes place during rapid cooling of the graphene-Cu system.

The model discussed so far considers single vacancies as the unique sources of entropy in the graphene system. The real situation is much more complicated, however the model can be extended to larger vacancies. At first, we can consider a di-vacancy as the result of the *chemical* reaction [21]: **$1V+1V=1V_2+ E_{2V}$**, where **$E_{2V}$** is the binding energy of the di-vacancy consequent to the collapse of two single vacancies. This means that:

$$H_{F(2V)} = 2H_{F(1V)} - E_{2V} . \tag{2.0}$$

It is trivial to demonstrate that the quantity **$E_{2V}$ ($E_{nV}$** in the case of a cluster with n-vacancies) adds to $\Delta$ in eq. (1.8), meaning that the *gain* in entropy when a cluster of n-vacancies is annihilated, is larger than in the case of n single vacancies.

Discontinuities in the underlying metal film are sources of large clusters of vacancies in the growing graphene layer. The case of grain boundaries is a well-known example. As it has been elucidated [11], the growth of graphene is impeded by grain boundaries in case of Cu (where precipitation is negligible), whereas enhanced growth of graphene is reported for the Ni and Co cases. The widely accepted explanation is the contribution to the growth from atoms precipitated from the vis-a-vis surfaces of two contiguous grains.

It has also been reported that graphene can grow on wide discontinuities of the metal (Ni, in that case) film [25]. This does not happen in the case of Cu: no Raman signatures of graphene are observed in holes opened by the dewetting phenomenon occurring in the Cu films before exposure to C precursor [26].

The occurrence of the growth of a suspended layer of graphene on holes in the underlying Co film deserves special consideration. The origin of such holes, few μm wide, is ascribable, like in case of Cu, to dewetting, which is unlikely in the case of pure Co (with melting temperature $T_m$=1495 °C), but can occur at C contents close to saturation ($T_m$ ~1300 °C [20]). Dewetting preferentially occurs at the boundary of three or more grains [27], and it is strongly inhibited by the presence of the growing graphene layer [12]. Basing on these considerations, it is likely that holes in the Co film form during the isothermal gas exposure step. In other words, the graphene growth occurs *On-Nothing* during the following step. Furthermore, the occurrence of the alternative mechanism, the growth of graphene at a grain boundary, followed by hole opening, would imply the formation of a multi-layer at that boundary. Raman investigation, on the contrary, confirms that the graphene on Co holes grows as defect-free monolayer.

Some considerations are needed about such growth of *Graphene-On-Nothing* in view of the present model. The graphene giant vacancy corresponds to the Co hole, and is a great source of entropy of the film. Nevertheless, the system tends to reduce its entropy by precipitation of C from the Co boundaries surrounding the hole, arranging them in order to progressively reduce entropy at the perimeter of the giant vacancy cluster. A continuous graphene layer is then obtained in spite of discontinuities in the metal substrate. This underlines the crucial role played by the high activation energy of C solubility in Co in sustaining this kind of unconventional growth.

**Conclusions**

In summary, a comparison among the different mechanisms of graphene growth in Cu, Ni, and Co, is presented. It is shown that C atoms precipitation from the metal substrate is an efficient mechanism to reduce the total entropy of the graphene layer. A model is proposed, which shows that not only the absolute value of C solubility, but also its dependence on temperature play an important role in reducing entropy. Thanks to the high activation energy of C solubility, Co shows advantages with respect to other, more employed, catalyzing metals.

This viewpoint is confirmed by the good structural and electrical properties of the graphene layers obtained in this way.

**Acknowledgments** The valuable discussions with Prof. E. Vittone, of the Department of Physics, University of Torino, are gratefully acknowledged. Work supported by the Istituto Nazionale di Ricerca Metrologica through the "Nanotechnologies for Electromagnetic Metrology" program.

**References**

1. Neumann, C., Reichardt, S., Venezuela, P., Drögeler, M., Banszerus, L., Schmitz, M., Watanabe, K., Taniguchi, T., Mauri, F., Beschoten, B., Rotkin, S.V., Stampfer, C., 2015. Raman spectroscopy as probe of nanometre-scale strain variations in graphene. Nat Commun 6, 8429. doi:10.1038/ncomms9429

2. Amato, G., Milano, G., Vignolo, U., Vittone, E., 2015. Kinetics of defect formation in chemically vapor deposited (CVD) graphene during laser irradiation: The case of Raman investigation. Nano Res. 8, 3972–3981. doi:10.1007/s12274-015-0900-1

3. Krauss, B., Lohmann, T., Chae, D.-H., Haluska, M., von Klitzing, K., Smet, J.H., 2009. Laser-induced disassembly of a graphene single crystal into a nanocrystalline network. Phys. Rev. B 79, 165428. doi:10.1103/PhysRevB.79.165428

4. R. McLellan, Scr. Metall., 1969, 3, 389.

5. Vilkov, O., Fedorov, A., Usachov, D., Yashina, L.V., Generalov, A.V., Borygina, K., Verbitskiy, N.I., Grüneis, A., Vyalikh, D.V., 2013. Controlled assembly of graphene-capped nickel, cobalt and iron silicides. Sci. Rep. 3. doi:10.1038/srep02168

6. Ago, H., Ito, Y., Mizuta, N., Yoshida, K., Hu, B., Orofeo, C.M., Tsuji, M., Ikeda, K., Mizuno, S., 2010. Epitaxial chemical vapor deposition growth of single-layer graphene over cobalt film crystallized on sapphire. ACS Nano 4, 7407–7414. doi:10.1021/nn102519b

7. Decker, R., Brede, J., Atodiresei, N., Caciuc, V., Blügel, S., Wiesendanger, R., 2013. Atomic-scale magnetism of cobalt-intercalated graphene. Phys. Rev. B 87, 041403. doi:10.1103/PhysRevB.87.041403

8. Li, X., Cai, W., Colombo, L., Ruoff, R.S., 2009. Evolution of Graphene Growth on Ni and Cu by Carbon Isotope Labeling. Nano Lett. 9, 4268–4272. doi:10.1021/nl902515k

9. Kalbac, M., Frank, O., Kavan, L., 2012. The control of graphene double-layer formation in copper-catalyzed chemical vapor deposition. Carbon 50, 3682–3687. doi:10.1016/j.carbon.2012.03.041

10. Wofford, J.M., Nie, S., McCarty, K.F., Bartelt, N.C., Dubon, O.D., 2010. Graphene Islands on Cu Foils: The Interplay between Shape, Orientation, and Defects. Nano Lett. 10, 4890–4896. doi:10.1021/nl102788f

11. Mattevi, C., Kim, H., Chhowalla, M., 2011. A review of chemical vapour deposition of graphene on copper. J. Mater. Chem. 21, 3324–3334. doi:10.1039/C0JM02126A

12. Croin, L., Vittone, E., Amato, G., 2014. In situ control of dewetting of Cu thin films in graphene chemical vapor deposition. Thin Solid Films 573, 122–127. doi:10.1016/j.tsf.2014.10.102

13. Cultrera, A., Amato, G., Boarino, L., Lamberti, C., 2014. A modified cryostat for photo-electrical characterization of porous materials in controlled atmosphere at very low gas dosage. AIP Advances 4, 087134. doi:10.1063/1.4894074

14. Bae, S., Kim, H., Lee, Y., Xu, X., Park, J.-S., Zheng, Y., Balakrishnan, J., Lei, T., Ri Kim, H., Song, Y.I., Kim, Y.-J., Kim, K.S., Özyilmaz, B., Ahn, J.-H., Hong, B.H., Iijima, S., 2010. Roll-to-roll production of 30-inch graphene films for transparent electrodes. Nat Nano 5, 574–578. doi:10.1038/nnano.2010.132

15. Lee, J.-K., Park, C.-S., Kim, H., 2014. Sheet resistance variation of graphene grown on annealed and mechanically polished Cu films. RSC Adv. 4, 62453–62456. doi:10.1039/C4RA11734D

16. Kwon, K.C., Ham, J., Kim, S., Lee, J.-L., Kim, S.Y., 2014. Eco-friendly graphene synthesis on Cu foil electroplated by reusing Cu etchants. Scientific Reports 4. doi:10.1038/srep04830

17. Kim, K.S., Zhao, Y., Jang, H., Lee, S.Y., Kim, J.M., Kim, K.S., Ahn, J.-H., Kim, P., Choi, J.-Y., Hong, B.H., 2009. Large-scale pattern growth of graphene films for stretchable transparent electrodes. Nature 457, 706–710. doi:10.1038/nature07719

18. Bruna, M., Cassiago, C., Callegaro, L., Gasparotto, E., Rocci, R., Borini, S., 2010. Fabrication and characterization of graphene-based quantum hall effect devices at INRIM, in: 2010 Conference on Precision Electromagnetic Measurements (CPEM). Presented at the 2010 Conference on Precision Electromagnetic Measurements (CPEM), pp. 349–350. doi:10.1109/CPEM.2010.5544271

19. Weatherup, R.S., Bayer, B.C., Blume, R., Baehtz, C., Kidambi, P.R., Fouquet, M., Wirth, C.T., Schlögl, R., Hofmann, S., 2012. On the Mechanisms of Ni-Catalysed Graphene Chemical Vapour Deposition. ChemPhysChem 13, 2544–2549. doi:10.1002/cphc.201101020

20. Hasebe, M., Ohtani, H., Nishizawa, T., 1985. Effect of magnetic transition on solubility of carbon in bcc Fe and fcc Co-Ni alloys. MTA 16, 913–921. doi:10.1007/BF02814843

21. W. Hayes and A. M. Stoneham: Defects and Defect Processes in Nonmetallic Solids, Dover Publications (2004), **ISBN:** 0486434834

22. Berry, R.S., Rice, S.A., Ross, J. *Physical Chemistry*, Part II. 1980; pp 564, Wiley, New York.

23. Ohtani, H., Hasebe, M., Nishizawa, T., 1984. Calculation of Fe-C, Co-C and Ni-C Phase Diagrams. Transactions of the Iron and Steel Institute of Japan 24, 857–864. doi:10.2355/isijinternational1966.24.857

24. Weatherup, R.S., Baehtz, C., Dlubak, B., Bayer, B.C., Kidambi, P.R., Blume, R., Schloegl, R., Hofmann, S., 2013. Introducing Carbon Diffusion Barriers for Uniform, High-Quality Graphene Growth from Solid Sources. Nano Lett. 13, 4624–4631. doi:10.1021/nl401601x

25. Wessely, P.J., Wessely, F., Birinci, E., Beckmann, K., Riedinger, B., Schwalke, U., 2012. Silicon-CMOS compatible in-situ CCVD grown graphene transistors with ultra-high on/off-current ratio. Physica E: Low-dimensional Systems and Nanostructures 44, 1132–1135. doi:10.1016/j.physe.2011.12.022

26. Croin, L., Vittone, E., Amato, G., 2014. Low-temperature rapid thermal CVD of nanocrystalline graphene on Cu thin films. Phys. Status Solidi B 251, 2515–2520. doi:10.1002/pssb.201451219

27. Ye, J., Thompson, C.V., 2011. Anisotropic edge retraction and hole growth during solid-state dewetting of single crystal nickel thin films. Acta Materialia 59, 582–589. doi:10.1016/j.actamat.2010.09.062

**Figure Captions**

Fig. 1. Low energy portion of a typical Raman spectrum of a graphene sample before (right) and after (left) transfer. The lower intensity of the spectrum on the right is due to the light absorption from the Co substrate.

Fig 2. Scanning Electron Micrograph of the electrical device: the metallic strips are 150 μm wide and 766 μm spaced. The overall area between the two internal contacts is about 1.32 mm$^2$.

Fig. 3. AFM (a) and SEM (b) images of the graphene layer grown on a micrometric hole originating from Co dewetting.

Fig. 4. Scanning Electron Micrograph of transferred graphene showing small bilayer islands detected as darker areas. The biggest one is evidenced by the dashes.

Fig. 5. Raman maps of a 50 μm X 50 μm portion of a transferred graphene sample. In a) the normalized intensity $I_{2D}/I_G$ is mapped, and the corresponding frequency histogram is reported in b). Same results are displayed in c) and d) for the $I_D/I_G$ ratio.

Fig. 6. Time evolution for the normalized intensity of the *D* peak in the case of graphene deposited on Cu (a), and Co (b).

Fig. 7. The dependence on the inverse temperature of the solubility (in wt. %) of C in Co (squares), Ni (circlets) and Cu (triangles). Data for Co and Ni are from Ref. [20], whereas the Cu ones have been manually extracted from the Cu-C phase diagram reported in Ref. [11].

Figure 1

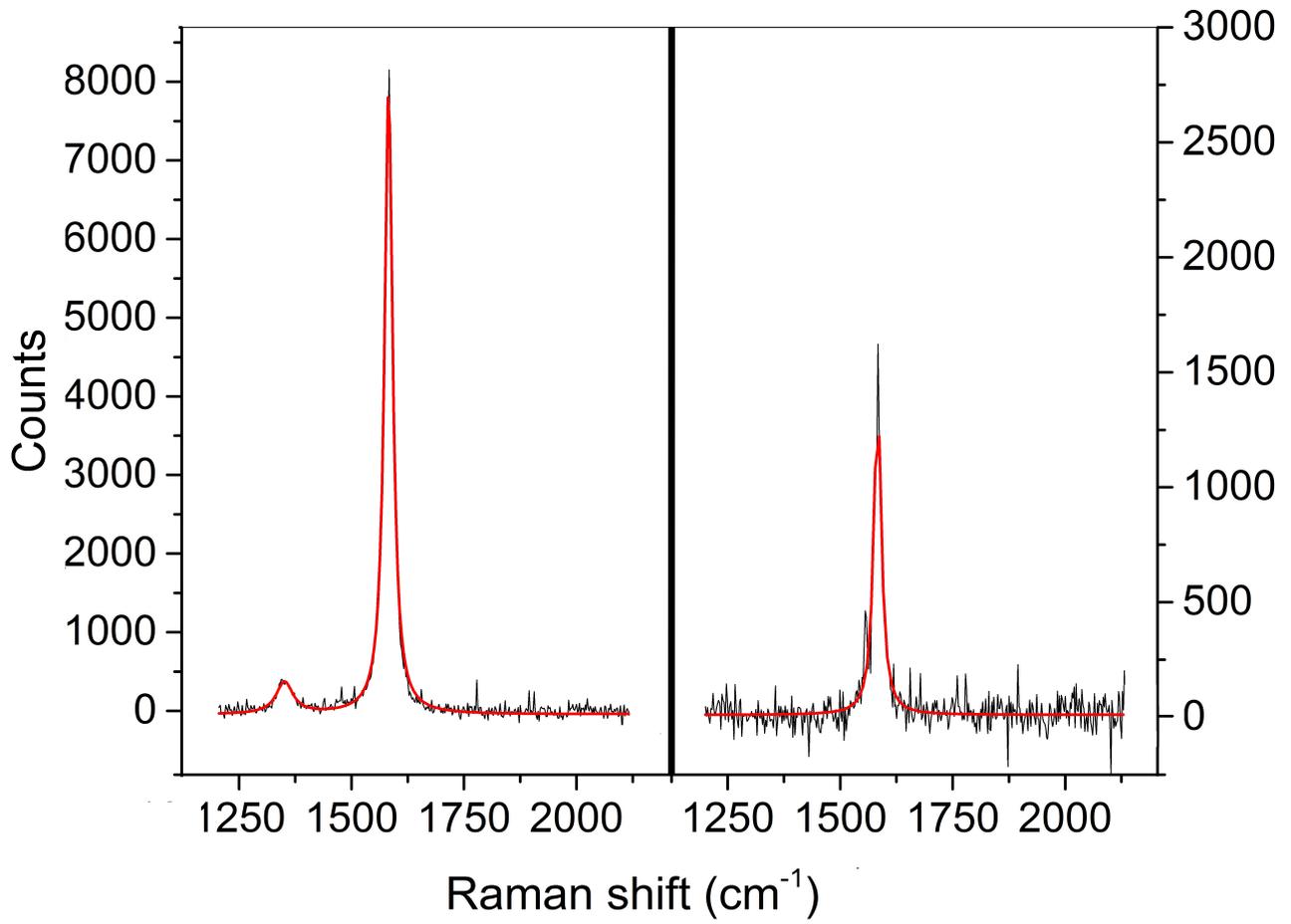

Figure 2

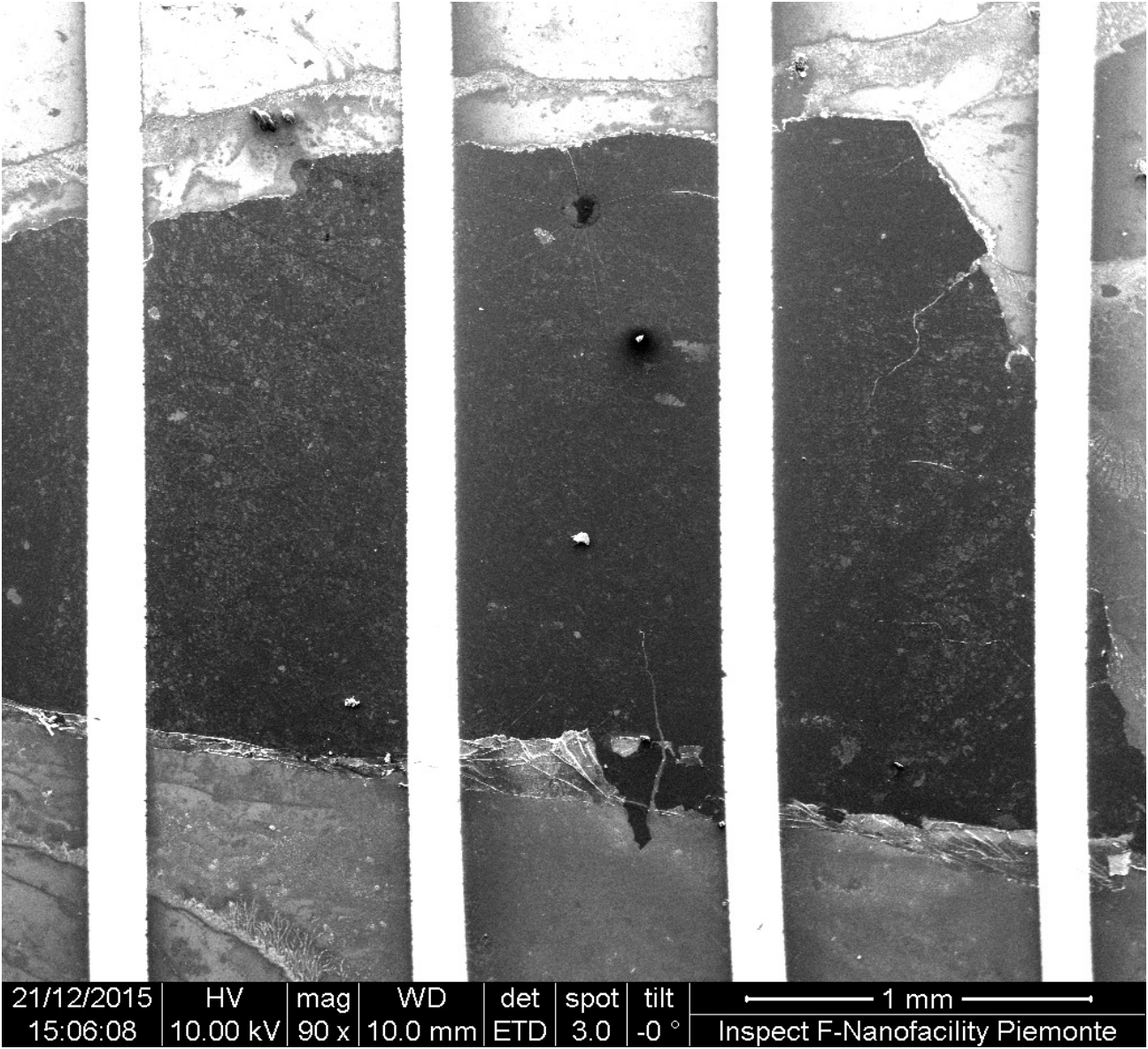

Figure 3a

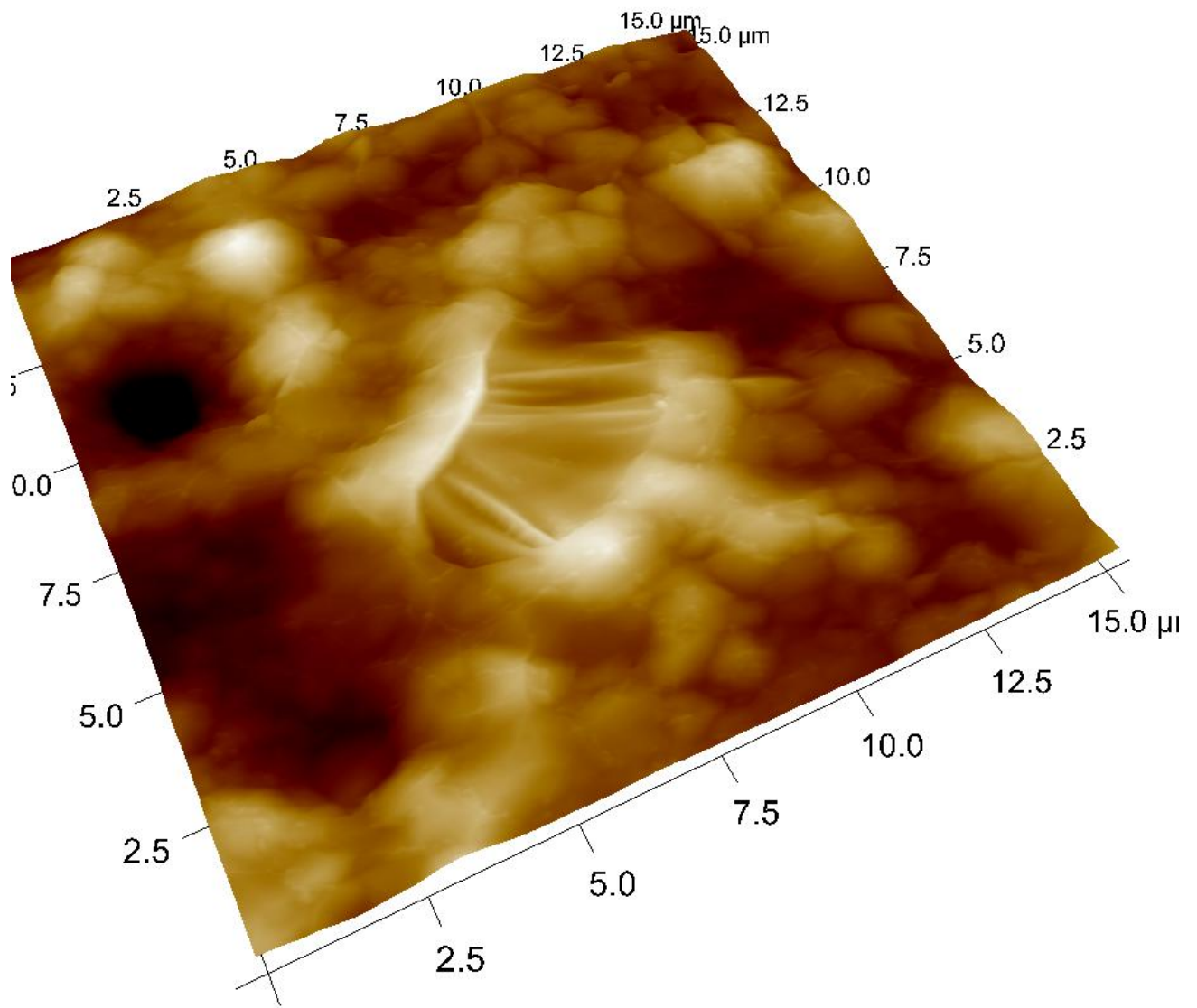

Figure 3b

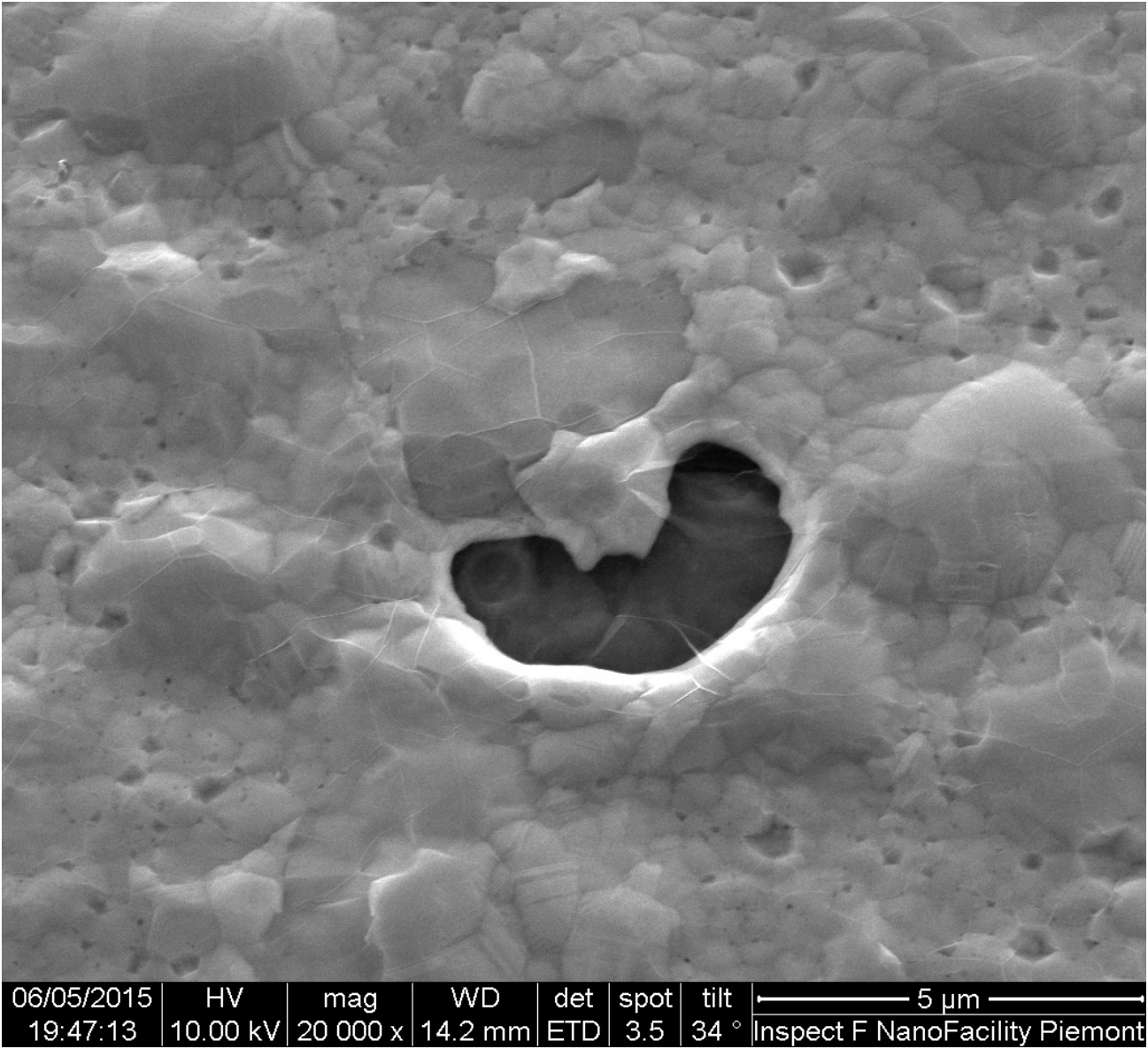

Figure 4

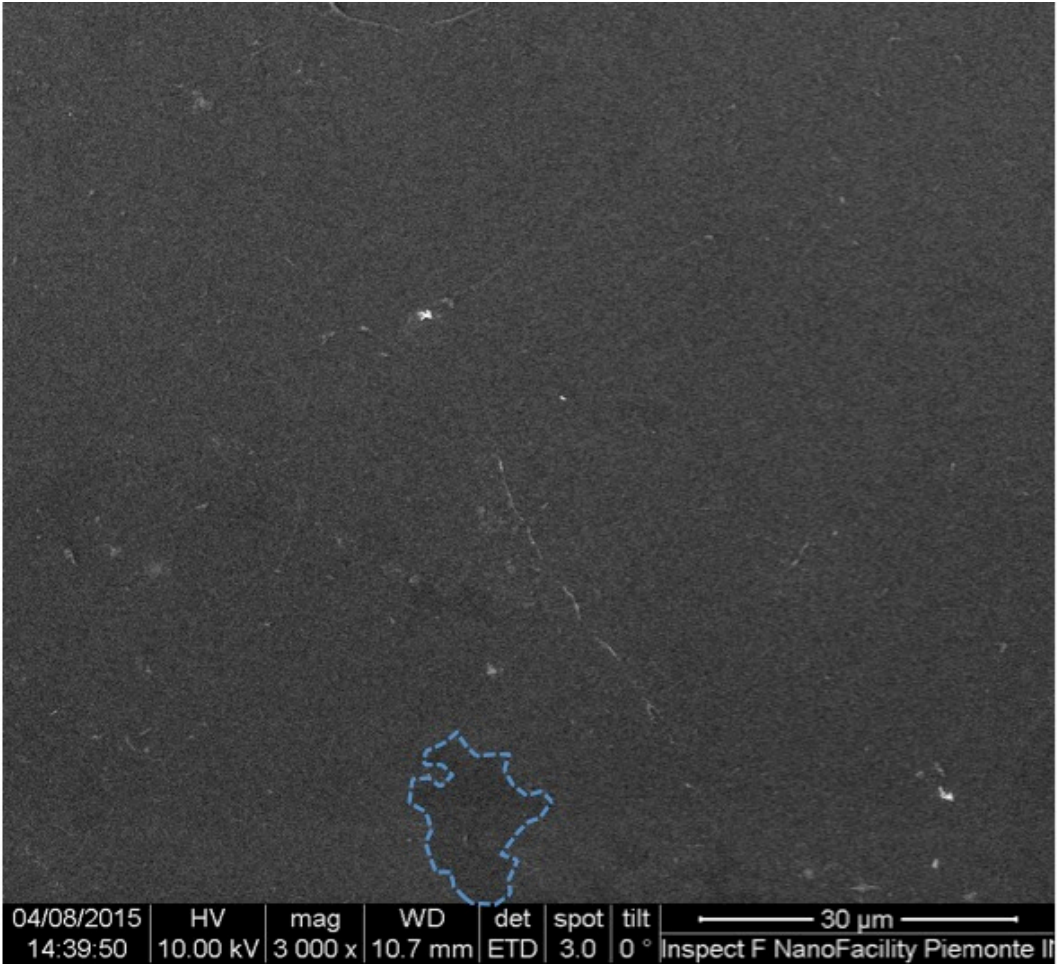

Figure 5a

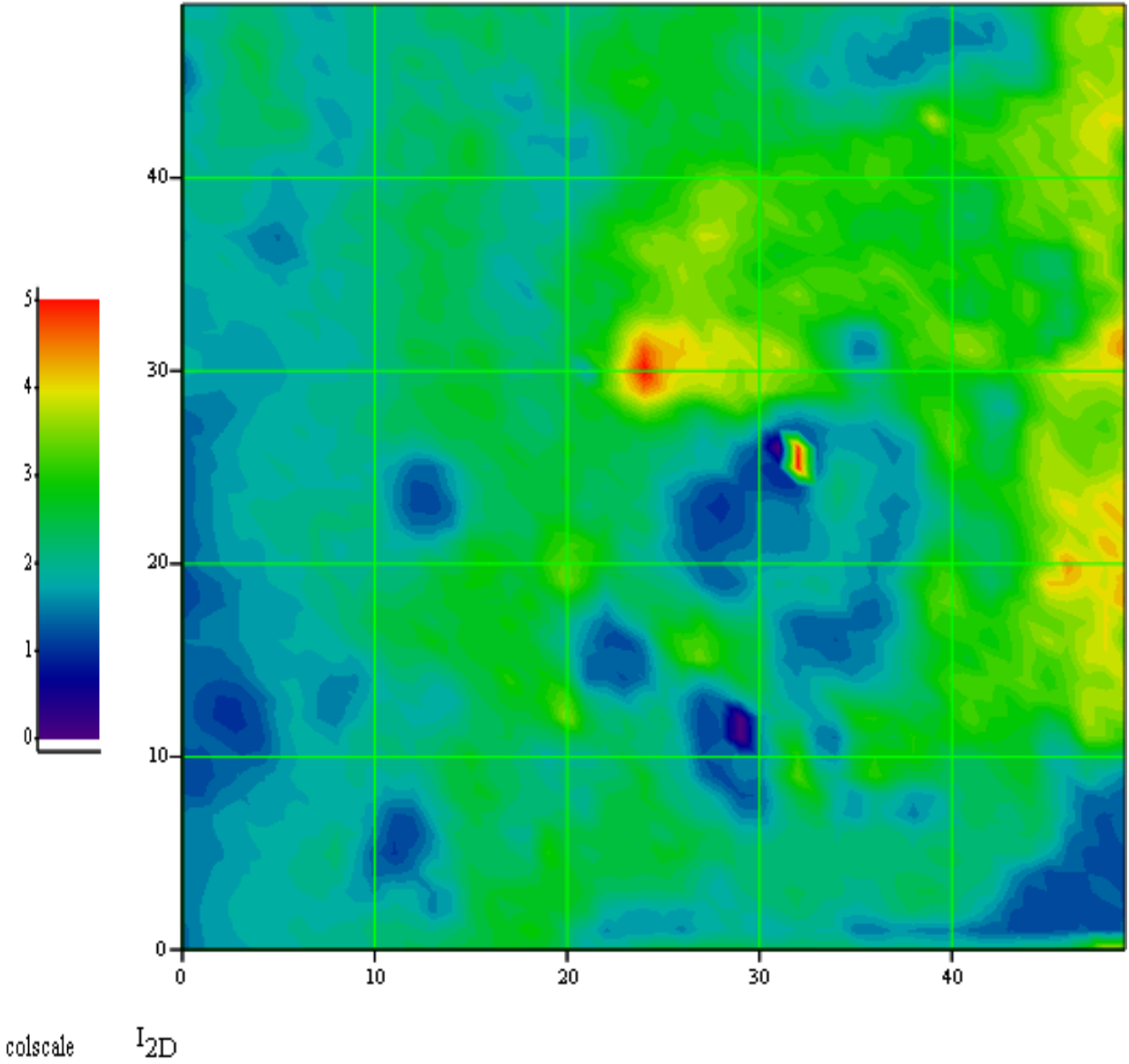

colscale   $I_{2D}$

Figure 5b

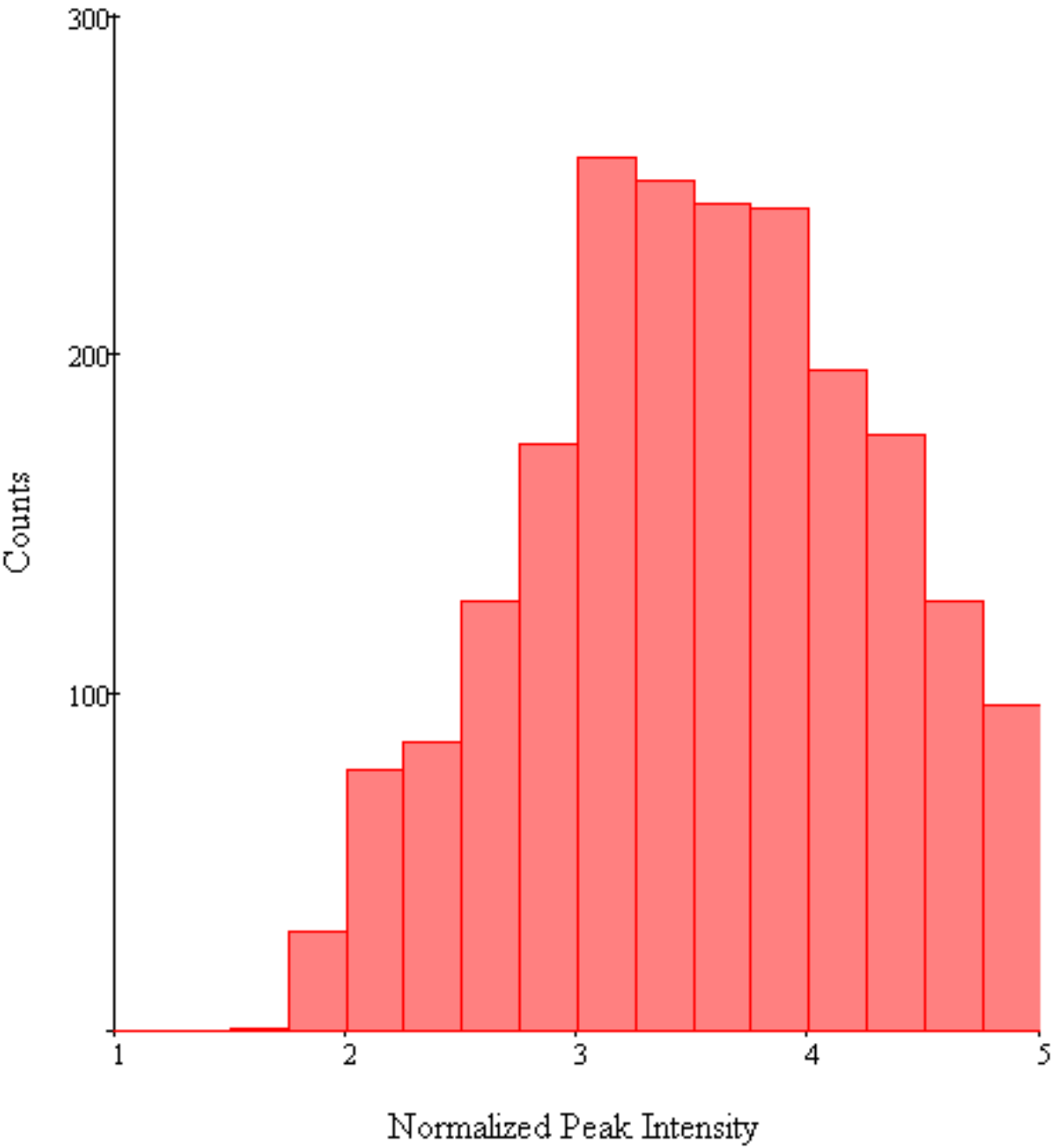

Figure 5c

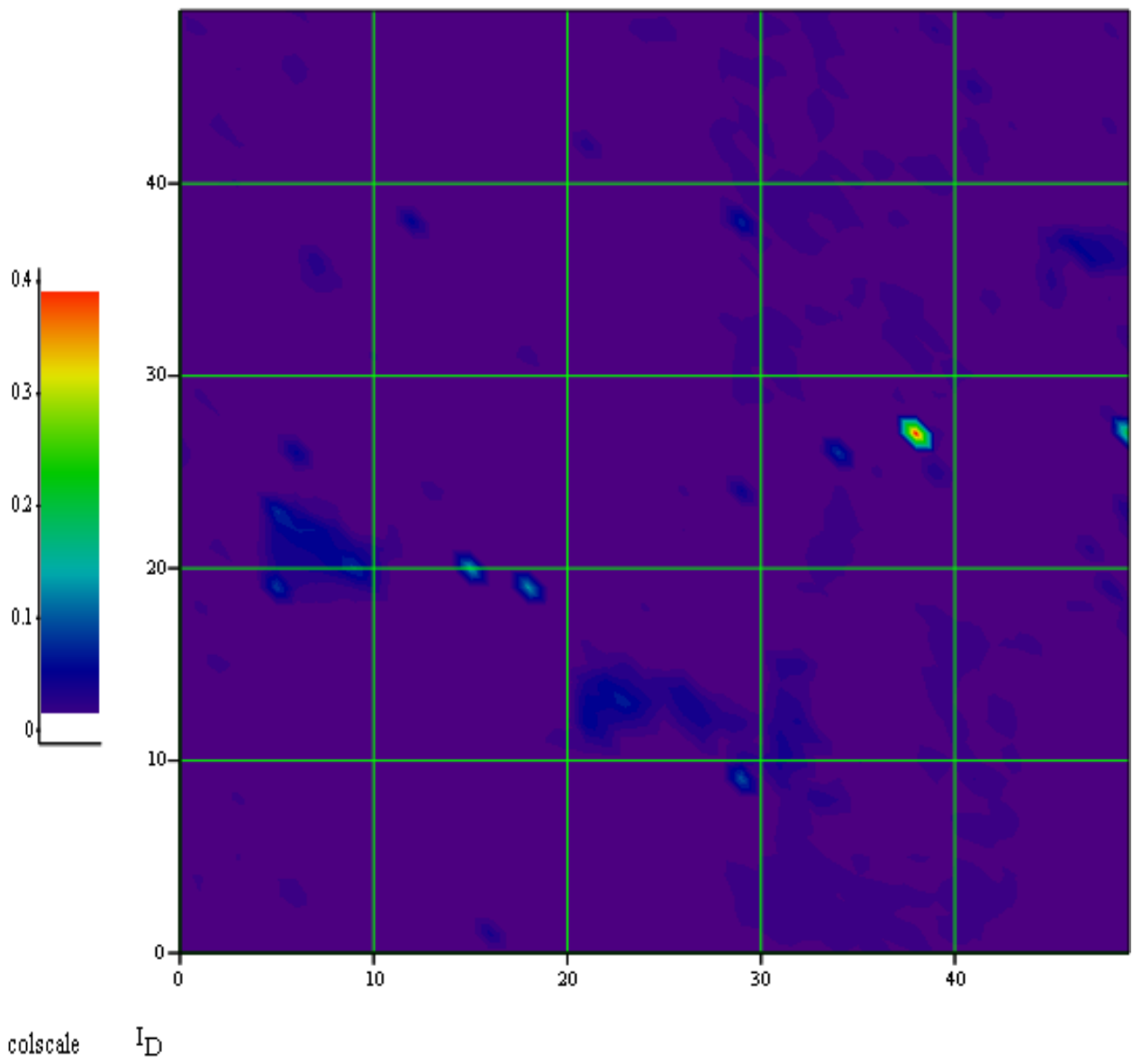

colscale    $I_D$

Figure 5d

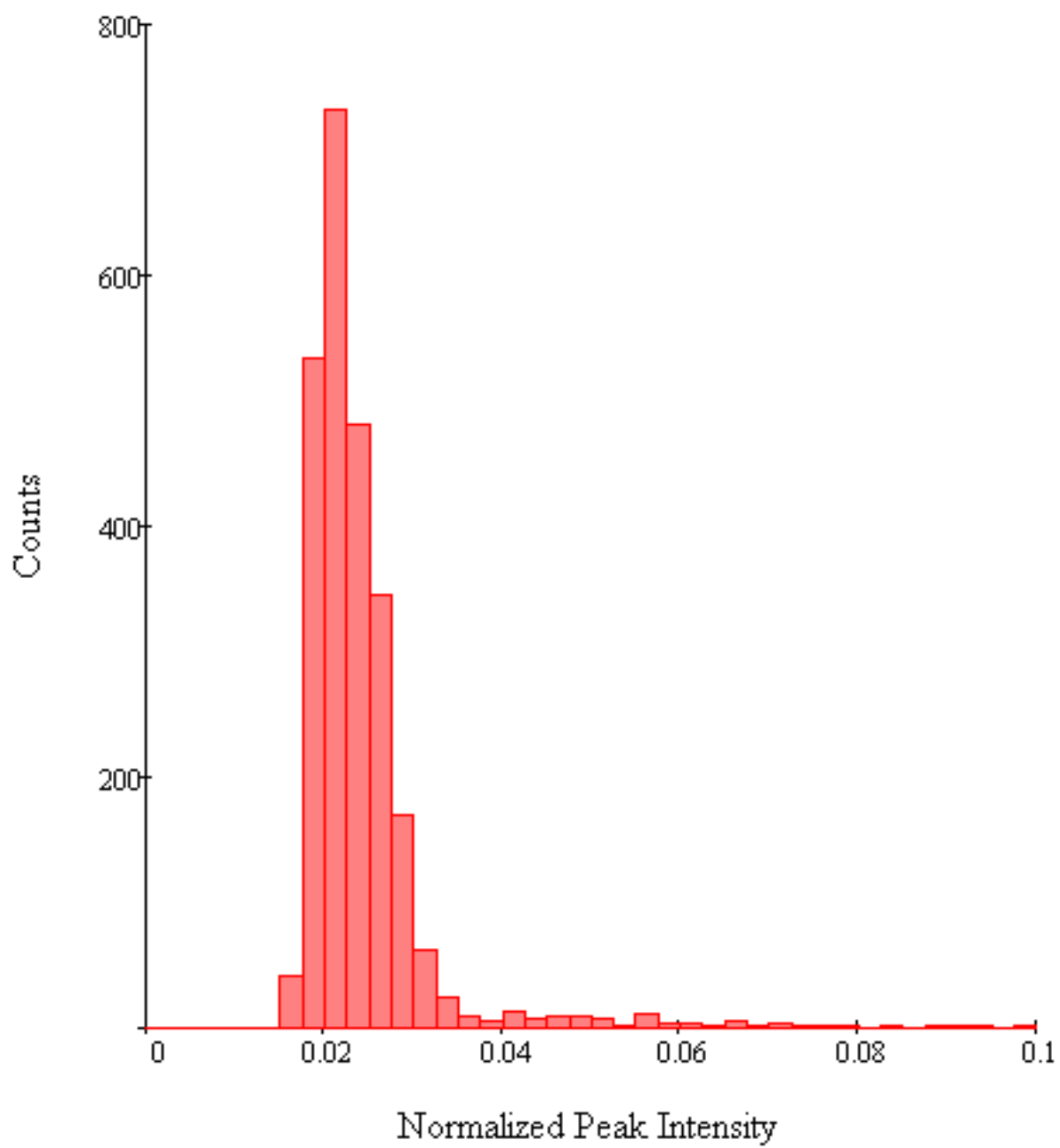

Figure 6a

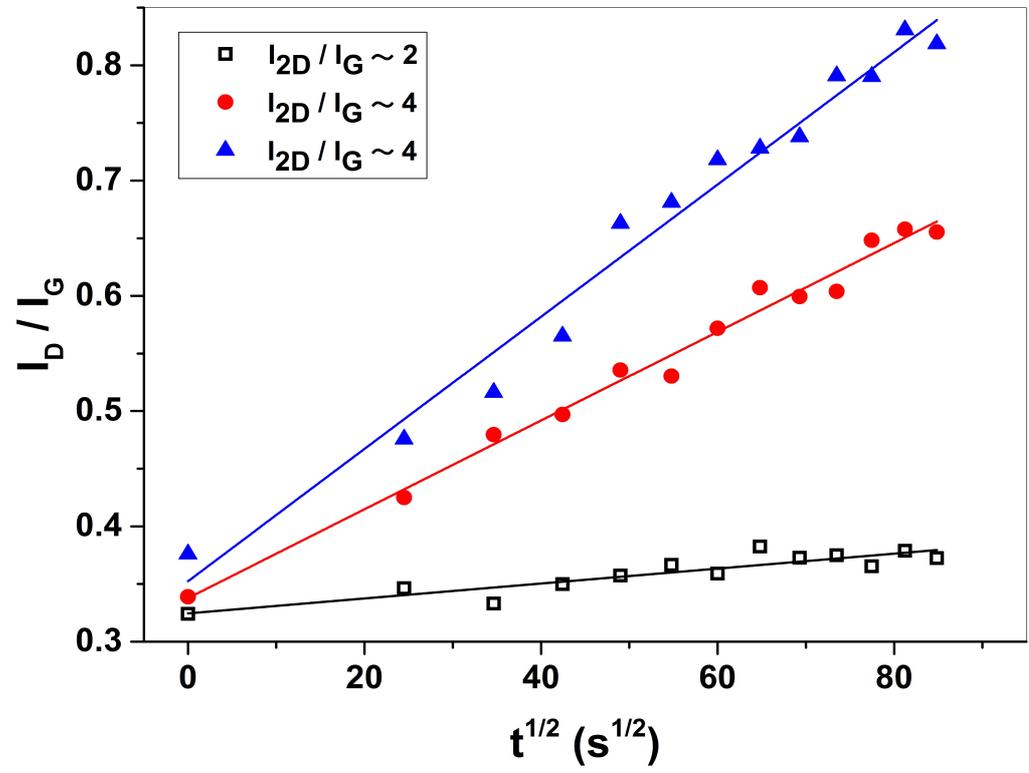

Figure 6b

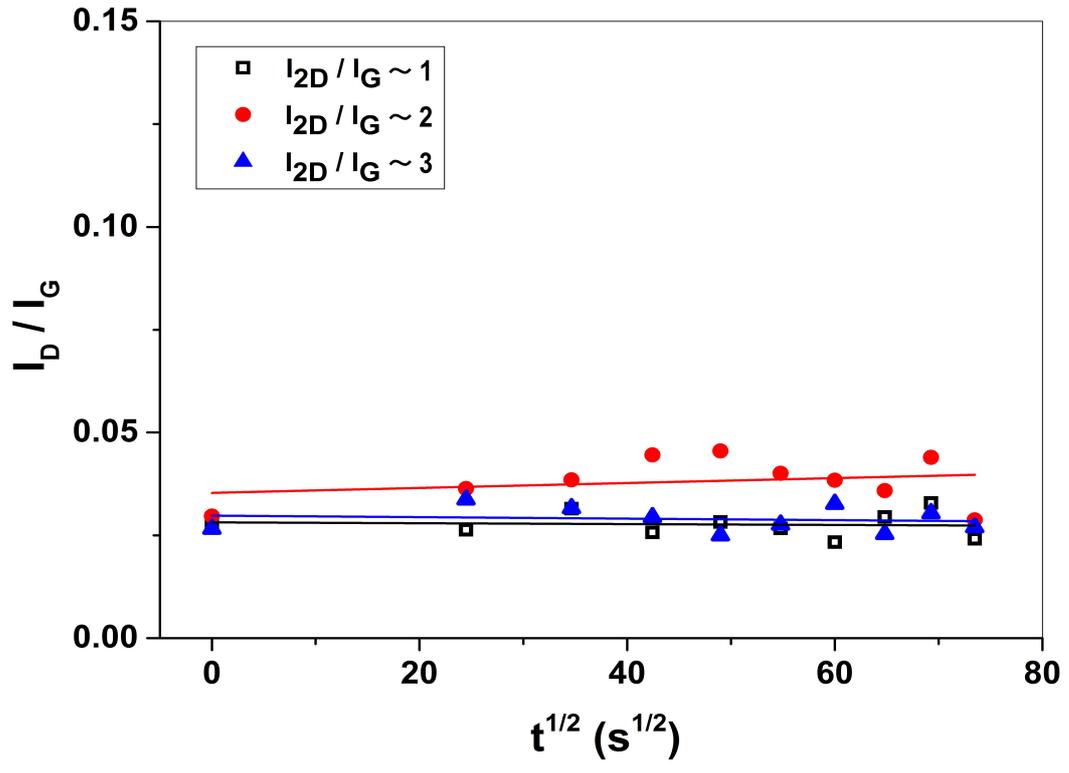

Figure 7

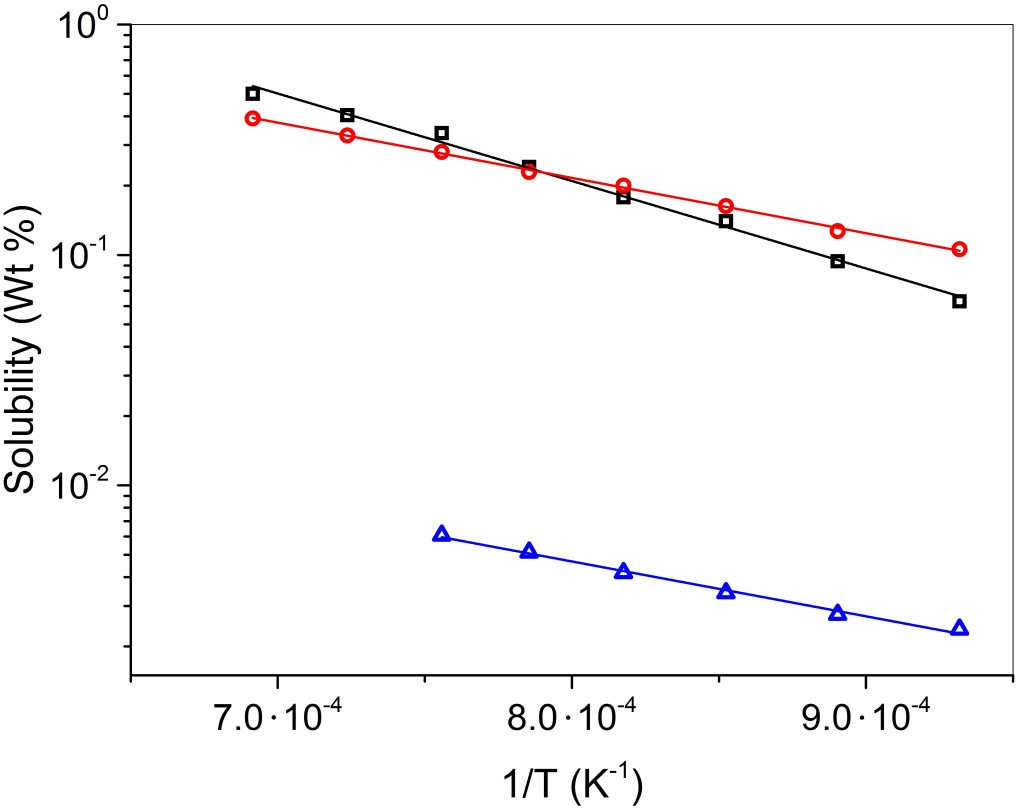